\pdfoutput=1
\documentclass[11pt]{article}
\usepackage{geometry}
\usepackage[sort&compress,numbers,colon]{natbib}
\usepackage{hyperref}

\usepackage{amsmath,amssymb,authblk}
\usepackage{graphicx}
\usepackage{tikz}
\usepackage{booktabs}
\usepackage{slashbox}
\usepackage{multirow}

\def\ket#1{\mathinner{|{#1}\rangle}}

\newcommand{\Br}{\ensuremath{\mathrm{Br}}}

\definecolor{MS}{rgb}{1,0,0}
\definecolor{YC}{rgb}{0,0,1}

\makeatletter

\newcommand{\Rmnum}[1]{\expandafter\@slowromancap\romannumeral #1@}
\makeatother

\usepackage{subcaption}

\usepackage{tikz}
\usetikzlibrary{positioning,arrows,patterns}
\usetikzlibrary{decorations.pathmorphing}
\usetikzlibrary{decorations.markings}
\usetikzlibrary{calc}
\tikzset{
    photon/.style={decorate, decoration={snake}, draw=black, thick},
    fermionnoarrow/.style={draw=black, postaction={decorate}, thick},
    scalar/.style={draw=black, postaction={decorate}, thick, dashed},
    fermion/.style={draw=black, postaction={decorate},decoration={markings,mark=at position .55 with {\arrow{>}}}, thick},
    gluon/.style={decorate, draw=black, decoration={coil,amplitude=4pt, segment length=5pt}, thick},
    vertex/.style={draw,shape=circle,fill=black,minimum size=3pt,inner sep=0pt}
}

\begin{document}
\title{A Case Study of the Sensitivity to LFV Operators with Precision Measurements and the LHC}
\author[1]{Yi Cai\thanks{\texttt{yi.cai@unimelb.edu.au}}}
\author[2]{Michael A. Schmidt\thanks{\texttt{michael.schmidt@sydney.edu.au}}}
\affil[1]{ARC Centre of Excellence for Particle Physics at the Terascale, School of Physics,
The University of Melbourne, Victoria 3010, Australia}
\affil[2]{ARC Centre of Excellence for Particle Physics at the Terascale, School of Physics, The University of Sydney, NSW 2006, Australia}
\date{}

\maketitle

\begin{abstract}
We compare the sensitivity of precision measurements of lepton flavour
observables to the reach of the LHC in a case study of lepton-flavour violating
operators of dimension six with two leptons and two quarks. For light quarks
precision measurements always yield the more stringent constraints. The LHC
complements precision measurements for operators with heavier quarks.
Competitive limits can already be set on the cutoff scale
$\Lambda>600-800$ GeV for operators with right-handed $\tau$ leptons using the LHC run 1 data.
\end{abstract}

\section{Introduction}
The discovery of the $125 \; \rm{GeV}$ Higgs boson~\cite{ATLAS:2012gk,CMS:2012gu} in 2012 at the Large Hadron Collider (LHC) has completed the   
description of the highly successful Standard Model (SM) of particle physics.
However, a number of experimental observations and theoretical arguments, such as the origin of neutrino masses, the existence of dark matter, the hierarchy problem and the strong CP problem, can not be accommodated within the SM.
Many theoretical proposals addressing these issues generally lead to lepton flavour violating (LFV) 
processes which are theoretically forbidden in the SM by accidental symmetries.  
The prime examples are models of neutrino mass. The observation of neutrino
oscillations~\cite{Fukuda:1998mi} undeniably showed that individual lepton
number is not conserved. Thus LFV processes, such as $\mu \rightarrow e \gamma$,
may exist. In the minimal type-I seesaw model~\cite{Minkowski:1977sc}, these processes are suppressed by unitarity and far out of current and future experimental reach. 
However, in the other two seesaw models~\cite{Cheng:1980qt, foot:1988aq} and also in radiative neutrino mass models~\cite{Zee:1980ai},
LFV processes enjoy more freedom and their rates can be large enough to be tested.   
Other examples include but are not limited to ($R$-parity violating) supersymmetric models~\cite{Barbier:2004ez} and $Z^\prime$ models~\cite{Langacker:2008yv}.

The observation of these LFV processes will definitely shed light on the deeper underlying physics,
while the non-observation surely places stringent constraints on the model parameters of the proposed theories.  
The classical experiments try to search for very rare processes such as $\mu^-\rightarrow e^-\gamma$, 
$\mu^-\rightarrow e^-e^+e^-$, $\mu$-$e$ conversion in nuclei and rare $\tau$ and LFV meson decays 
at MEG~\cite{Adam:2013mnn}, Mu3e~\cite{Blondel:2013ia}, Mu2E~\cite{Carey:2008zz,Kutschke:2011ux}, COMET~\cite{Hungerford:2009zz,Cui:2009zz}, SINDRUM~\cite{Dohmen:1993mp}, B-factories~\cite{Aubert:2001tu, Abashian:2000cg}, {\it et al}. 
We will refer to these experiments as {\it precision measurements} due to the ultra-high experimental sensitivities.  
Meanwhile, LFV processes can also occur at collider experiments with a relatively low SM background.
For example, in supersymmetric models squarks and gluinos can be produced at the Tevatron or the LHC with subsequent LFV decays in
a cascade decay chain via sleptons. Of this type of collider tests, we will focus on {\it the LHC} since 
the results are generally the best ones.      

So far such LFV processes have not been observed from precision measurements.
At the LHC, the CMS experiment recently reported a 2.4 $\sigma$ anomaly in the $h\to \mu\tau$ decay~\cite{Khachatryan:2015kon},
while the analysis of ATLAS~\cite{Aad:2015gha} is consistent with the SM and the CMS result.   
All these experimental results suggested that the energy scale $\Lambda$ 
where new physics emerges are rather high and much larger than the electroweak scale. 
Therefore we can adopt a simple formalism to  interpret the experimental results, namely the effective operators. 

In light of the LHC particularly interesting operators are the ones with two
quarks and two leptons, because they allow for relatively large cross sections
and clean signatures with low SM background. There are ten different gauge
invariant operators with two quarks and two leptons, denoted by representations
of $SU(2)_L$, following the discussion in
Ref.~\cite{Buchmuller:1985jz,Grzadkowski:2010es}. After electroweak symmetry
breaking, the gauge-invariant operators induce different contributions to the
four-fermion interactions of neutrinos, charged leptons and quarks, which directly enter the relevant physical processes. Constraints obtained for the individual four-fermion interactions can be translated to constraints on the gauge-invariant effective operators by using the most stringent constraint of the generated four-fermion interactions of quarks and leptons.
We consider the $SU(2)_L$ invariant operators, obtain the corresponding
effective four fermion interactions and determine the most stringent constraints
both from precision experiments and the LHC.
Previous studies~\cite{Carpentier:2010ue,Petrov:2013vka} of effective operators with two quarks and two leptons focused on constraints from precision experiments and did not aim to explore the potential of the LHC.

The paper is organised as follows: in Sec.~\ref{sec:operators} we discuss the
LFV effective operators and choose one type for our case study. Although we
restricted ourselves to one operator, operator mixing will induce other
operators. We discuss QCD renormalization group (RG) corrections in
Sec.~\ref{sec:RG}.
Then we study the constraints on the chosen operator from precision measurements in Sec.~\ref{sec:flavour}.
In Sec.~\ref{sec:lhc}, we recast the relevant study on the LFV processes from the LHC and draw the current limits and also the future projection at Run 2. We summarise and discuss our results in Sec.~\ref{sec:discussion}.
Sec.~\ref{sec:con} is devoted to the conclusion. Technical details are collected
in the appendix.

\section{Effective Operators}
\label{sec:operators}
Following the general classification of dimension six
operators~\cite{Buchmuller:1985jz,Grzadkowski:2010es}, there are
10 dimension six operators with two quark and two lepton fields neglecting the
flavour structure 
\begin{align}
	\mathcal{Q}_{lq}^{(1)}&= (\bar L\gamma_\mu L) (\bar Q\gamma^\mu Q)\;,&
	\mathcal{Q}_{lq}^{(3)} &= (\bar L\gamma_\mu \tau^I L)(\bar Q \gamma^\mu
\tau^I Q)\;, \\
\mathcal{Q}_{eu} &=(\bar \ell \gamma_\mu \ell)(\bar u \gamma^\mu u)\;,&
\mathcal{Q}_{ed} &=(\bar \ell \gamma_\mu \ell)(\bar d \gamma^\mu d)\;,\\
\mathcal{Q}_{lu} &= (\bar L \gamma_\mu L)(\bar u\gamma^\mu u)\;,&
\mathcal{Q}_{ld} &= (\bar L \gamma_\mu L)(\bar d\gamma^\mu d)\;,&
\mathcal{Q}_{qe} &= (\bar Q \gamma_\mu Q)(\bar \ell\gamma^\mu \ell)\;,\\
\label{eq:SMeffOp}
\mathcal{Q}_{ledq}&= (\bar L^\alpha \ell)(\bar d Q^\alpha)\;,&
\mathcal{Q}_{lequ}^{(1)}&=(\bar L^\alpha \ell)\epsilon_{\alpha\beta}(\bar
Q^\beta u)\;,\\
\mathcal{Q}_{lequ}^{(3)}&=(\bar L^\alpha\sigma_{\mu\nu}
\ell)\epsilon_{\alpha\beta}(\bar Q^\beta \sigma^{\mu\nu}u)\;,
\end{align}
where $\alpha$, and $\beta$ are SU(2)$_L$ indices.

\begin{figure}[tb]
\centering
\begin{tikzpicture}[node distance=1cm and 1cm]

\coordinate[vertex] (v);
\coordinate[vertex,above left = of v] (v1);
\coordinate[vertex,below left = of v] (v2);

\coordinate[left = of v1, label=left:$g$] (g1);
\coordinate[left = of v2, label=left:$g$] (g2);
\coordinate[above right = of v, label=right:$\ell_i$] (l1);
\coordinate[below right = of v, label=right:$\ell_j$] (l2);

\draw[fermionnoarrow] (v1) -- (v2) node[midway,label=left:$t$] {};
\draw[fermionnoarrow] (v2) -- (v) node[midway,label=right:$t$] {};
\draw[fermionnoarrow] (v) -- (v1) node[midway,label=right:$t$] {};

\draw[gluon] (v1) to (g1);
\draw[gluon] (v2) to (g2);

\draw[fermionnoarrow] (v) -- (l2);
\draw[fermionnoarrow] (v) -- (l1);
\end{tikzpicture}

\begin{minipage}{10cm}
\caption{Feynman diagram for the operator with top-quark bilinear to generate LFV final states at the LHC. \label{fig:top}}

\end{minipage}
\end{figure}
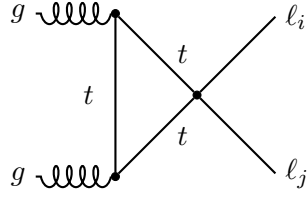

In general, the quark bilinears can be any combination of flavours and the
leptonic bilinear has to be flavour off-diagonal to explain LFV.
Various combinations of quark flavours will involve different mesons in the analysis.
To cover the whole spectra of mesons is definitely a mission that can not be contained in this single work.   
Thus we will only start with quark bilinears of same flavours, where we expect
the weakest constraints from precision experiments.
Among those, the operator with the top quark pair bilinear can only contribute
at one-loop at the LHC as shown in Fig.~\ref{fig:top}, which leads to an
effective dimension 7 operator at low energies with two gluons field strength
tensors coupled to a lepton bilinear.
This operator has completely different flavour constraints from other operators with lighter quark bilinears.
Thus we will restrict ourselves to a study of effective operators with the first five flavour quarks and leave the operator with top quarks for future study.    

Both $t$-channel scalar exchange and $s$-channel vector boson exchange generate
operators with vector bilinears, which have been studied intensively in
Ref.~\cite{Carpentier:2010ue,Petrov:2013vka} in terms of effective four fermion
interactions. Thus, we will take the operators generated via an $s$-channel
scalar exchange $\mathcal{Q}_{ledq}$ and $\mathcal{Q}_{lequ}^{(1)}$ in
Eq.~\eqref{eq:SMeffOp} with Wilson coefficients $\Xi^{d}$ and $\Xi^u$,
respectively, as a
fresh example to demonstrate our study of the sensitivity with precision
measurements and the LHC
\begin{equation}\label{eqn:op}
	-\mathcal{L}=\Xi_{ij,kl}^{d}
	\left(\mathcal{Q}_{ledq}\right)_{ij,kl}+\Xi_{ij,kl}^{u}
	\left(\mathcal{Q}_{lequ}^{(1)}\right)_{ij,kl} + \mathrm{h.c.}\;.
\end{equation}
 They, for instance, are generated in two Higgs doublet models with tree-level flavour violation~\cite{Bjorken:1977vt,hep-ph/9601383v1,1401.6147v2,Branco:2011iw}. We will, however, be agnostic about the underlying UV completion and will study the effective operators without any theoretical prejudice.

Specifically, we exemplify the possibility to test effective operators
$\mathcal{Q}_{ledq}$, $\mathcal{Q}_{lequ}^{(1)}$ with two leptons $\ell_{i,j}$ and two same-flavour quarks $q_k$ besides the top quark, $t$, at the LHC and in precision experiments.  It is straightforward to extend the study to operators with a different Lorentz structure.

Writing the SU(2)$_L$ structure explicitly, the operators read 
\begin{align}
	\left(\mathcal{Q}_{ledq}\right)_{ij,kl}&= (\bar L_i^\alpha \ell_j)(\bar d_k Q^\alpha_l) = 
	(\bar \nu_{Li} \ell_{Rj}) (\bar d_{Rk} u_{Ll}) + 
	(\bar \ell_{Li} \ell_{Rj}) (\bar d_{Rk} d_{Ll}) \;,\\
	\left(\mathcal{Q}_{lequ}^{(1)}\right)_{ij,kl}&=(\bar L_i^\alpha \ell_j)\epsilon_{\alpha\beta}(\bar Q_k^\beta u_l)=
	(\bar \nu_{Li} \ell_{Rj}) (\bar d_{Lk} u_{Rl}) - 
	(\bar \ell_{Li} \ell_{Rj}) (\bar u_{Lk} u_{Rl}) 
\end{align}
and thus lead to two effective four fermion interactions. We define the Wilson coefficients of the effective four fermion
interactions $\Xi^{Nu}$, $\Xi^{Nd}$, $\Xi^{Cu}$, $\Xi^{Cd}$ as follows
\begin{align} 
	\mathcal{L}_{4f}=&
	\Xi^{Cd}_{ij,kl} (\bar \nu_{Li} \ell_{Rj}) (\bar d_{Rk} u_{Ll}) +
	\Xi^{Nd}_{ij,kl} (\bar \ell_{Li} \ell_{Rj}) (\bar d_{Rk} d_{Ll})
 \\\nonumber &
	+\Xi^{Cu}_{ij,kl} (\bar \nu_{Li} \ell_{Rj}) (\bar d_{Lk}
u_{Rl}) +\Xi^{Nu}_{ij,kl} (\bar \ell_{Li} \ell_{Rj}) (\bar u_{Lk} u_{Rl}) \;.
\end{align}
They are related to the Wilson coefficients in the unbroken theory via
\begin{align}
	\Xi^{Nd}_{ij,kl}&= U^{\ell*}_{ii^\prime}\, V^d_{ll^\prime}
	\,\Xi^d_{i^\prime j,kl^\prime}\;,&
	\Xi^{Cd}_{ij,kl}&= U^{\nu*}_{ii^\prime}\, V^u_{ll^\prime}\, \Xi^d_{i^\prime
j,kl^\prime}\;,\\
\Xi^{Nu}_{ij,kl}&=-U^{\ell*}_{ii^\prime} \,V^{u*}_{kk^\prime}
\,\Xi^u_{i^\prime j,k^\prime l}\;, &
\Xi^{Cu}_{ij,kl}&=U^{\nu*}_{ii^\prime}\, V^{d*}_{kk^\prime}\, \Xi^u_{i^\prime
j,k^\prime l}\;,
\end{align} 
where the unitary matrices $U^{\ell,\nu}$ and $V^{u,d}$ relate the quark and lepton
states in the basis where the dimension six operator is defined to their mass
eigenstates denoted by subscript $m$, i.e.
\begin{align} 
	\nu &= U^\nu \nu_m\;,& 
	\ell &= U^\ell \ell_m\;,& 
	u &= V^d u_m\;,&
	d &= V^d d_m\;.
\end{align} 
In the following discussion we choose the charged leptons to be diagonal, i.e.
$U^\ell=\mathbf{1}$ and thus $U^\nu$ becomes the PMNS matrix $U$. Furthermore, we
choose the Wilson coefficients $\Xi^{Nu,Nd}$ to be diagonal in the quark
sector. This choice implies that there are no flavour changing neutral currents at tree level.
In case of operator 
$\mathcal{Q}_{ledq}$, this implies $V^u=\mathbf{1}$ and $V^d$ becomes the CKM
matrix $V$. Similarly for operator 
$\mathcal{Q}_{lequ}^{(1)}$ we find $V^d=\mathbf{1}$ and $V^u=V^\dagger$.
We use the current best-fit values from the UTfit collaboration~\cite{Bona:2006ah} for
the CKM matrix and the ones of the nu-fit
collaboration~\cite{Gonzalez-Garcia:2014bfa} for the PMNS matrix assuming that
all leptonic CP phases vanish.

Generally, however, those operators are accompanied by operators with neutral
current quark-flavour-violating (QFV) operators. We will also quote limits from these induced operators. In particular, we parameterize the Wilson coefficients of the accompanying QFV operators by (no summation on the right-hand side)
\begin{align}\label{eq:opMix}
\Xi_{ij,kl}^{u}& = \lambda\, \Xi_{ij,ll}^{u} V_{kl}\;,&
\Xi_{ij,kl}^{d}& = \lambda\, \Xi_{ij,kk}^{d} V_{kl}
\end{align} 
for up-type and down-type quarks, where $\lambda$ indicates the mixing induced from matching to the full theory, which is normalised to the corresponding CKM mixing matrix element.

All Wilson coefficients are fixed at the scale $\mu=1$ TeV. Thus in order to make connection with results from low-energy precision
experiments, we have to include RG corrections.

\section{Renormalization Group Running}
\label{sec:RG}
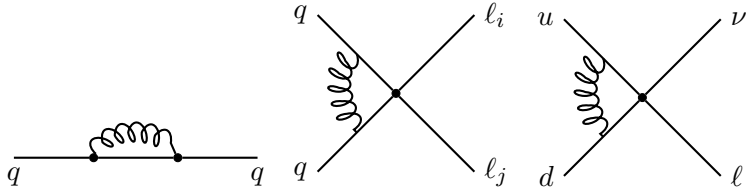
\begin{figure}[b]\centering
\begin{tikzpicture}[node distance=1cm and 1cm]

\coordinate[vertex] (v1);
\coordinate[vertex,right = of v1] (v2);
\coordinate[left = of v1, label=below:$q$] (f1);
\coordinate[right = of v2, label=below:$q$] (f2);

\draw[fermionnoarrow] (f1) -- (v1);
\draw[fermionnoarrow] (f2) -- (v2);

\draw[fermionnoarrow] (v1) -- (v2);

\draw[gluon,out=60] (v1) to[in=120] (v2);
\end{tikzpicture}
\begin{tikzpicture}[node distance=1cm and 1cm]
\coordinate[vertex] (v);
\coordinate[above left = of v, label=left:$q$] (f1);
\coordinate[below left = of v, label=left:$q$] (f2);
\coordinate[above right = of v, label=right:$\ell_i$] (f3);
\coordinate[below right = of v, label=right:$\ell_j$] (f4);

\draw[fermionnoarrow] (f1) -- (v);
\draw[fermionnoarrow] (f2) -- (v);
\draw[fermionnoarrow] (f3) -- (v);
\draw[fermionnoarrow] (f4) -- (v);

\draw[gluon,out=210] ($(f1)!0.5!(v)$) to[in=120] ($(f2)!0.5!(v)$);
\end{tikzpicture}
\begin{tikzpicture}[node distance=1cm and 1cm]
\coordinate[vertex] (v);
\coordinate[above left = of v, label=left:$u$] (f1);
\coordinate[below left = of v, label=left:$d$] (f2);
\coordinate[above right = of v, label=right:$\nu$] (f3);
\coordinate[below right = of v, label=right:$\ell$] (f4);

\draw[fermionnoarrow] (f1) -- (v);
\draw[fermionnoarrow] (f2) -- (v);
\draw[fermionnoarrow] (f3) -- (v);
\draw[fermionnoarrow] (f4) -- (v);

\draw[gluon,out=210] ($(f1)!0.5!(v)$) to[in=120] ($(f2)!0.5!(v)$);
\end{tikzpicture}
\caption{Relevant diagrams for QCD corrections, where $q$ collectively stands
for up- and down- quarks.}
\label{fig:QCDrunning}
\end{figure}
Particularly QCD corrections to the operators are important due to the size of
the strong interactions. We follow the discussion in Ref.~\cite{Buchalla:1995vs}
to include QCD corrections at next-to leading order to the operators. We take
into account the mass thresholds of the quarks and match the effective theories
with $n_F$ quark flavours at the pole mass of each quark. As there is no operator mixing for the two quark-two lepton operators from QCD running, the
next-to leading order QCD correction simplifies tremendously and the Wilson
coefficients at a scale $\mu$ are related to the ones at a scale $\mu_0$ via
\begin{align}
	\Xi(\mu)
&=\Xi(\mu_0)
\left(\frac{\alpha_s(\mu)}{\alpha_s(\mu_0)}\right)^{\frac{\gamma_0}{2\beta_0}}\;.
\end{align}
The relevant coefficients $\beta_0$ and $\gamma_0$
can be directly read from the beta functions of $\Xi$
\begin{align}
	\frac{d \Xi}{d\ln\mu} &
	= - \gamma_0 \frac{\alpha_s}{4\pi}\Xi
\end{align}
and the one-loop beta function of the strong coupling
\begin{equation}
	\frac{d \alpha_s}{d\ln\mu} = -2\beta_0 \frac{\alpha_s^2}{4\pi}\;.
\end{equation}
A straightforward calculation shows 
\begin{align}
	\beta_0=11-2 n_F/3\quad \mathrm{and}\quad
	\gamma_0 =6 C_2(3)\;,
\end{align}
where $C_2(3)=4/3$ is the quadratic Casimir invariant of the fundamental
representation. The relevant diagrams of QCD corrections to the quark propagator and the
effective vertex are shown in Fig.~\ref{fig:QCDrunning}.

We use the Mathematica code RunDec~\cite{Chetyrkin:2000yt} to obtain the strong coupling at the
different mass scales, which we use to evaluate the running of the Wilson
coefficients. Given the large uncertainty of the LHC analysis, we only use the two
loop QCD RG equations and match the effective theories at one-loop.

\section{Existing Flavour Physics Constraints}
\label{sec:flavour}
There are already several constraints on the operators in Eq.~\eqref{eqn:op} with same-flavour quarks from existing flavour experiments. The main constraints are from $\mu$-$e$ conversion, LFV neutral meson decays, leptonic charged pseudoscalar decays and semi-leptonic $\tau$-decays.

We do not take into account the recent hints for new physics in different $B$-decays measured at LHCb~\cite{Aaij:2013aln,Aaij:2013qta,Aaij:2014ora} or the recent hint for lepton flavour non-universality in $B\to D^*\tau\nu$ measured by BaBar~\cite{Lees:2012xj}, Belle~\cite{Huschle:2015rga}, and LHCb~\cite{Aaij:2029609}. An explanation of these hints for new physics requires operators with quark flavour violation. See Ref.~\cite{Calibbi:2015kma} for a recent study using effective operators.

We define the Wilson coefficients $\Xi^{u,d}_{ij,kk}$ with an arbitrary phase at the scale $\mu=1$ TeV
and evolve them down to the scale of the relevant process, like the mass of the
$\tau$ lepton or the heavy quark of the decaying meson. For $\mu$-$e$ conversion
and decays of light mesons with masses below $1$ GeV, we evaluate the operator at $\mu=1$ GeV, where the operators are matched to chiral perturbation theory, but we neglect any additional quantum corrections in chiral perturbation theory for simplicity.
The masses, decay constants, and mixing angles of the considered mesons are
summarised in App.~\ref{app:mesons}. Unless stated otherwise we use the
experimental values reported in Ref.~\cite{Agashe:2014kda}. We vary the phase of
the Wilson coefficient in steps of $1^\circ$ and report the range of obtained limits. We
quote all limits in terms of the cutoff scale $\Lambda$ of the effective
operators, i.e.
\begin{equation}
	\Lambda\equiv \Xi^{-1/2}\;.
\end{equation}

\mathversion{bold}
\subsection{$\mu$-$e$ conversion}
\mathversion{normal}
The conversion of $\mu$-$e$  in nuclei is probed for several different nuclei, like gold (Au), titanium (Ti), and lead (Pb).
So far no observation of the process has been made.  This places a stringent limit on the dimensionless 
$\mu$-$e$ conversion rate defined as,
\begin{equation}
R_{\mu e}^{\left(A,Z\right)}\equiv \frac{\Gamma\left(\mu^-+(A,Z)\to e^- +(A,Z)\right)}{\Gamma\left(\mu^-+(A,Z)\to \nu_\mu+(A,Z-1)\right)}\; ,
\label{eqn:emuconversion}
\end{equation}
where $A$ and $Z$ are the mass number and the atomic number of the nuclei.  
The denominator of Eq.~\eqref{eqn:emuconversion} denotes the well-measured muon capture rate and the numerator is the muon conversion rate calculated with
\begin{equation}
\Gamma\left(\mu^-+(A,Z)\to e^-+(A,Z)\right) =\left|\Xi^{Nu,Nd}_{ij,kk}\right|^2\times \mathcal{F} \times \frac{ p_e E_e \left(\mathcal{M}_p +\mathcal{M}_n\right)^2 }{2\pi} \; ,
\end{equation} 
where $p_e$ and $E_e$ is the momentum and energy of the final electron, 
and $\mathcal{M}_{p,n}$ are the nuclear matrix elements.
We follow Ref.~\cite{Gonzalez:2013rea} in the analysis and list both the muon total capture rates and the nuclear matrix elements in Tab.~\ref{tab:mueConstraints}.
The factor $\mathcal{F}$ parameterizes the interaction between the charged lepton current and the nuclei,
\begin{equation}
\mathcal{F}= \left|\alpha_{SS}^{(0)} +\alpha_{SS}^{(3)} \frac{\mathcal{M}_p-\mathcal{M}_n}{\mathcal{M}_p+\mathcal{M}_n} \right|^2 
+ \left|\alpha_{PS}^{(0)} +\alpha_{PS}^{(3)} \frac{\mathcal{M}_p-\mathcal{M}_n}{\mathcal{M}_p+\mathcal{M}_n} \right|^2 \; , 
\end{equation}
where parameters with superscripts $0$ and $3$ are related to isospin singlet and triplet respectively.
It can be described with two methods, {\it i.e.}~direct nuclear mediation and meson exchange mediation. Currently the relative strength of the two mechanisms is not known and for simplicity we separately consider them to obtain a limit and we expect that the actual limit will lie in between.

The direct nuclear mediation describes the interaction at the quark level, while the meson exchange mediation 
uses meson fields formed from the quark bilinear to mediate the interaction between the charged lepton and the nuclei.  
With direct nuclear mediation, the parameters $\alpha_{SS,PS}^{(0,3)}$ are given by
\begin{align}
 \alpha_{rS}^{(0,3)} = \frac{1}{4}\times \eta_{rS}^q\times 
 \begin{cases}
              G_S^{(0)}, G_S^{(3)} &  q = u\\
              G_S^{(0)}, -G_S^{(3)}&  q = d\\
              G_S^q, 0   & q=c, s, b  
 \end{cases}
\; ,
\end{align} 
where $r=S,P$, $G_S^{(0,3)}=(G_S^u \pm G_S^d)/2$, and 
the factor of $1/4$ in $\alpha_{rS}^{(0,3)}$ 
comes from the two projection operators in the quark bilinear and the lepton bilinear.
$\eta_{rS}^q$ takes $-1$ for $r=P$ with $\Gamma_l=P_L$ and takes $1$ in all other cases.    
The nucleon form factors take the following values~\cite{hep-ph/0507033v2,hep-ph/0107292v1}
\begin{equation}
G_S^u=3.74, 
\qquad G_S^d=2.694 ,
\qquad G_S^c=0.06,
\qquad G_S^s=0.64,
\qquad G_S^b=0.02 \; .
\end{equation}
Note, however, that there is significant uncertainty and the values might be up
to a factor of 2-4 larger as other calculations suggest~\cite{Gonzalez:2013rea}.

\begin{table}[tb]\centering
\begin{tabular}{lccc}
\toprule
& $^{48}$Ti & $^{197}$Au & $^{208}$Pb \\
\midrule
$p_e/\mathrm{fm}^{-1}$ & 0.529 & 0.485 & 0.482 \\
$\mathcal{M}_p/ \mathrm{fm}^{-3/2}$ & 0.104 & 0.395 & 0.414 \\
$\mathcal{M}_n/ \mathrm{fm}^{-3/2}$ & 0.127 & 0.516 & 0.566 \\
$\Gamma(\mu^-N\to \nu_\mu N) / 10^6 s^{-1}$ & 2.60 & 13.07 & 13.45 \\ 
$R_{\mu e}^\mathrm{max}$& $4.3\times 10^{-11}$\cite{Dohmen:1993mp} & $7.0\times 10^{-13}$\cite{Honecker:1996zf} & $4.6\times 10^{-11}$\cite{Bertl:2006up} \\
\midrule
\midrule
$\bar uu$   & $1100$ [$870$] & $2100$ [$1700$] & $760$ [$610$]\\
$\bar dd$   & $1100$ [$930$] & $2200$ [$1900$] & $780$ [$680$]\\
$\bar ss$ &  $480$ [-] & $950$  [-] & $340$ [-]\\
$\bar cc$ &  $150$ [-] & $290$  [-] & $110$ [-]\\
$\bar bb$ & $84$ [-] & $170$ [-]& $61$ [-]\\
\bottomrule
\end{tabular}
\caption{
Parameters for calculation of $\mu$-$e$ conversion rate 
and the constraints  on the cutoff scale $\Lambda$ [TeV] from $\mu$-$e$ conversion in nuclei using direct nuclear mediation [meson exchange mediation]. We obtain the same constraints for right-handed and left-handed operators. Similarly the constraints are symmetric in the leptons and it does not depend on the lepton bilinear $\bar\mu P_L e $ vs. $\bar e P_L \mu$.}
\label{tab:mueConstraints}
\end{table}

With the meson mediation method, the lepton bilinear will couple to an intermediate meson which also couples to the nuclei.    
Because of the Lorentz structure of the effective operators considered in this work, the only relevant mesons 
scalar mesons isosinglet $f_0(500)$ and isotriplet $a_0(980)$. The relevant parameters are
\begin{align}
\alpha_{rS}^{(0,3)} = \frac{1}{4} \times \eta_{rS}^q 
\begin{cases}
 \beta_{f_0} ,  \beta_{a_0}  & q =u \\ 
 \beta_{f_0} , - \beta_{a_0}  & q =d \\ 
 0                          & q = c, s, b
\end{cases} 
\; ,
\end{align}
where the parameters are estimated to be $\beta_{f_0}=1.58$ and $\beta_{a_0}=2.24$ as in~\cite{Gonzalez:2013rea}.
  
The current best limits on the conversion in these nuclei are $R_{\mu\nu}\leq 4.3\times 10^{-11}, 4.6\times 10^{-11}, 7.0\times 10^{-13}$ in $^{48}$Ti\cite{Dohmen:1993mp},  $^{208}$Pb\cite{Honecker:1996zf}, and $^{197}$Au\cite{Bertl:2006up}. Following Ref.~\cite{Gonzalez:2013rea} we calculate the constraint for the different quark flavours and summarise the results in Tab.~\ref{tab:Constraints}. Given the experimental constraints, $\mu$-$e$ conversion in gold leads to the most stringent constraint on the cutoff scale $\Lambda$. Direct nuclear mediation generally gives stronger constraints, particularly for the heavier quarks. If it would be entirely described by meson exchange mediation, the effective operators with heavier quarks are not constrained, because the form factors of all considered mesons vanish.
 
\mathversion{bold}
\subsection{Semi-Leptonic $\tau$-Decays}
\mathversion{normal}
Semi-leptonic $\tau$-decays impose another important constraint on operators with $\tau$ leptons and light quarks.
For the operators considered in this work, the only relevant and well-measured
$\tau$-decay modes are decays to pseudoscalar mesons $\pi^0$, $\eta$, $\eta^\prime$ and $K_S^0$ and to the scalar meson $f_0(980)$ which subsequently decays to pions.
We list the kinematically allowed channels and the limit on the branching ratios in Tab.~\ref{tab:tauDecays},
where we quote the current experimental limit on the branching ratios~\cite{Agashe:2014kda}.

\begin{table}[b!]
\centering
\resizebox{\linewidth}{!}{
	\begin{tabular}{lc ccc}
\toprule
decay & $\Br_i^{max}$ & \multicolumn{3}{c}{cutoff scale $\Lambda$ [TeV]}\\[2ex]
      & & $\Xi^{u}_{ij,uu}$ & $\Xi^{d}_{ij,dd}$   & $\Xi^{d}_{ij,ss}$  \\
\midrule
$\tau^- \to e^-\pi^0 $  &  $8.0\times 10^{-8}$ & $10$ & $10$ &  -\\
$\tau^- \to e^-\eta $ &  $9.2\times 10^{-8}$ & $34$ & $34$& $7.9$  \\
$\tau^- \to e^- \eta^\prime $ &     $ 1.6\times 10^{-7}$ &$42$ & $42$& $12$ \\
$\tau^- \to e^- K_S^0 $ &  $2.6\times 10^{-8}$  & -   & $7.8\,\sqrt{\lambda}$ &
$7.8\,\sqrt{\lambda}$ \\
$\tau^- \to e^- ( f_0(980)\to\pi^+\pi^-)$ &  $3.2\times 10^{-8}$   &$
13\,\sqrt{\sin\varphi_m}$  & $13\,\sqrt{\sin\varphi_m}$ & $16\, \sqrt{\cos\varphi_m}$\\
\midrule
$\tau^- \to \mu^-\pi^0 $ & $1.1\times 10^{-7}$ & $9.0-9.6$  & $9.0-9.6$& -\\
$\tau^- \to  \mu^-\eta$& $6.5\times 10^{-8}$ & $36-38$ & $36-38$& $8.4-8.9$  \\
$\tau^- \to  \mu^-\eta^\prime$&  $ 1.3\times 10^{-7}$ & $42-46$ &$42-46$ & $12-13$\\
$\tau^- \to  \mu^- K_S^0$&  $2.3\times 10^{-8}$  & - & $(7.8-8.3)\,\sqrt{\lambda}$ &
$(7.8-8.3)\,\sqrt{\lambda}$ \\
$\tau^- \to \mu^- ( f_0(980)\to\pi^+\pi^-)$&  $ 3.4\times 10^{-8}$
	 &$(12-14)\,\sqrt{\sin\varphi_m}$ &$(12-14)\,\sqrt{\sin\varphi_m}$ & $(15-16) \sqrt{\cos\varphi_m}$ \\
\bottomrule
\end{tabular}
}
\caption{Semi-leptonic $\tau$-decays.  Experimental constraint on the cutoff scale $\Lambda$ [TeV] of the effective operators. $\lambda$ denotes the mixing angle inducing operator mixing as defined in Eq.~\eqref{eq:opMix} and $\varphi_m$ is the mixing angle between $f_0(500)$ and $f_0(980)$ and is defined in Eq.~\eqref{eq:f0Mixing}.}
\label{tab:tauDecays}
\end{table}

The decay width for a $\tau^+$-lepton to a lighter lepton $\ell^+$ with mass $m_\ell$ and a neutral meson $M^0_{kk}=(\bar q_k q_k)$ is given by  
\begin{equation}
\Gamma(\tau^+\rightarrow \ell^+ M_{kk}^0)= \frac{k_M}{32\pi} \frac{m_M^2 \bar f_M^2}{m_{\tau}^2}
 \left[\left(m_\tau^2+m_\ell^2-m_M^2\right)|\Xi_\pm |^2 + 2 m_{\tau}m_\ell \mathrm{Re}\left(\Xi_\pm^2\right)\right]\;,
\end{equation} 
where $k_M$ is the magnitude of the meson 3-momentum in the centre-of-momentum frame
\begin{equation}
k_M^2=\frac{(m_\tau^2-(m_\ell+m_M)^2)(m_\tau^2-(m_\ell-m_M)^2)}{4m_\tau^2}
\end{equation}
and the effective coupling $\Xi_\pm$ is defined as
\begin{equation}\label{eq:Xidef}
\Xi_\pm\equiv\Xi^{Nu}_{ij,kl}\, \cos\varphi \pm\Xi^{Nd}_{ij,kl}\,\sin\varphi\;.
\end{equation}
$\Xi_+$ is the coupling for a scalar meson and $\Xi_-$ is the coupling for a pseudo-scalar meson in the final state.
The up-type (down-type) quark content of the meson is $\cos\varphi$ ($\sin\varphi$). 
The scale-dependent scalar meson decay constant $\bar f_M$ is defined in Eq.~\eqref{eq:decayconstants}.
 The partial decay width will be compared with the total decay width 
$
\Gamma_\tau =  \tau_\tau^{-1} = 2.27\times 10^{-9} \; \rm{MeV}\;.
$

Besides the pseudoscalar mesons, we consider the scalar meson $f_0(980)$, which  dominantly decays to two pions with a branching ratio Br$(f_0(980)\rightarrow \pi^+\pi^-) =0.46$ \cite{Aaij:2013zpt}. We parameterize its quark content by the mixing angle $\varphi_m$, which is defined in Eq.~\eqref{eq:f0Mixing}.

Our limits are quoted in Tab.~\ref{tab:Constraints}. 
The result only very weakly depends on the phase of the of the Wilson
coefficient $\Xi^{Nu,Nd}_{ij,kl}$ for hierarchical lepton masses and generally leads to a correction at the level of 
\begin{equation}
\frac{4m_{\ell_i}m_{\ell_j}}{m_{\ell_i}^2+m_{\ell_j}^2} \sim 4 \frac{\min \{m_{\ell_k}\}}{\max \{m_{\ell_k}\}}
\end{equation}
percent compared to the total decay width, which amounts to about 1\% (10\%) in case of an electron (muon) final state. Thus it is below the precision for an electron in the final state, but we quote the range in case of a muon in the final state.
The strongest limits are from decays to $\eta^{(\prime)}$ and $f_0(980)$ mesons
because the product $m_M \bar f_M$ is relatively large.  

\subsection{Leptonic Neutral Meson Decays}

Another important class of constraints comes from LFV neutral meson decays.
The decay width of a meson $M_{kl}^0=(\bar q_k q_l)$ can be expressed as
\begin{equation}
\Gamma({M^0_{kl}}\rightarrow \ell_i \ell_j)= \frac{k_\ell}{16\pi} \bar f_M^2   
 \left[\left(m_M^2-m_{\ell_i}^2-m_{\ell_j}^2\right)|\Xi_- |^2 + 2 m_{\ell_i} m_{\ell_j} \mathrm{Re}\left(\Xi_-^2\right)\right]\;,
\end{equation}
where $k_\ell$ is the magnitude of the lepton 3-momentum in the centre-of-momentum frame,
\begin{equation}\label{eq:kl}
k_\ell^2=\frac{(m_M^2-(m_{\ell_i}+m_{\ell_j})^2 )(m_M^2-(m_{\ell_i}-m_{\ell_j})^2)}{4m_M^2}
\end{equation}
and the  effective coupling $\Xi_-$ is defined in Eq.~\eqref{eq:Xidef}. The
experimental constraints on the cutoff scale $\Lambda$ [TeV] of the effective
operators are collected in Tab.~\ref{tab:LFVprocesses}. The top part of the
table lists the direct constraints on the operators with the same quarks in
Eq.~\eqref{eqn:op}, while the lower part summarises indirect constraints on the
operators from operator mixing induced by their creation from gauge invariant
operators. These constraints are parameterised by $\lambda$, which is defined in Eq.~\eqref{eq:opMix}. It is clear that we can place the strongest limit on operators with $e\mu$.

\begin{table}[btp]
\centering
\begin{tabular}{lc cccccc}
\toprule
decay &  $\Br_i^{max}$ & \multicolumn{5}{c}{cutoff scale $\Lambda$ [TeV]}\\[2ex]
 & & $\Xi^{u}_{ij,uu}$   &
$\Xi^{d}_{ij,dd}$  &
$\Xi^{d}_{ij,ss}$  & $\Xi^{u}_{ij,cc}$   &
$\Xi^{d}_{ij,bb}$  \\
\midrule
 $\pi^0\rightarrow \mu^+ e^-$  & $3.8\times 10^{-10}$ 
			       &$2.2$ & $ 2.2$ &-&-&-\\
 $\pi^0\rightarrow \mu^- e^+$  & $3.4\times 10^{-9}$ 
			       & $1.2 $ & $1.2 $ & -&-&-\\
 $\pi^0\rightarrow \mu^+ e^- + \mu^- e^+$  & $3.6\times 10^{-10}$ 
					   & $2.6$ &$2.6$  &-&-&-\\
 $ \eta\rightarrow \mu^+ e^- + \mu^- e^+$  & $6\times 10^{-6}$ 
					   & $0.52$ & $0.52 $ & $0.12 $  &-&-\\
 $\eta^\prime\rightarrow e\mu$  & $4.7\times 10^{-4}$ 
				& $0.091 $  & $ 0.091$ & $0.026$ &-&-\\
\midrule
\midrule
$K_L^0\to e^\pm \mu^\mp$ & $4.7\times 10^{-12}$ 
			 &-& $86\, \sqrt{\lambda}$ &$86\,\sqrt{\lambda}$ &-&-\\
 $D^0\to e^\pm \mu^\mp$ & $2.6\times 10^{-7}$ 
			& $6.4\,\sqrt{\lambda}$ &- &- & $6.4\,\sqrt{\lambda}$&-\\
$B^0\to e^\pm \mu^\mp$ & $2.8\times 10^{-9}$  
		       &-& $10\,\sqrt{\lambda}$ &- &-& $6.6\, \sqrt{\lambda}$\\
 $B^0\to e^\pm \tau^\mp$ & $2.8\times 10^{-5}$ 
			 &-&$0.97\,\sqrt{\lambda}$ &- &-& $0.62\,\sqrt{\lambda}$\\
 $B^0\to \mu^\pm \tau^\mp$ & $2.2\times 10^{-2}$ 
			   &-& $0.18\,\sqrt{\lambda}$ &- &-& $0.12\,\sqrt{\lambda}$\\
\bottomrule
\end{tabular}
\caption{Leptonic LFV  meson decays. Experimental constraint on the cutoff scale $\Lambda$ [TeV] of the effective operators. The processes listed in the top part of the table directly constrain the operators with the same quarks in Eq.~\eqref{eqn:op}, while the ones in the lower part indirectly constrain the operators with the same quarks via the operators generated by operator mixing as defined in Eq.~\eqref{eq:opMix}.}
\label{tab:LFVprocesses}
\end{table}

\subsection{Leptonic Charged Meson Decays}
As discussed in Sec.~\ref{sec:operators} there are also effective four fermion
interactions which contribute to different charged meson decays. 
Many charged meson decays have already been  measured and can be used to indirectly constrain the operators in Eq.~\eqref{eqn:op}. 
In particular the decays of $\pi^+$ and $K^+$ have been measured to high precision,
\begin{align}
R_\pi&= \frac{\Br(\pi^+\to e^+ \nu)}{\Br(\pi^+\to\mu^+\nu)}= \left(1.230\pm
0.004\right)\times 10^{-4}\;,& 
	\Br(\pi^+\to\mu^+\nu)&=0.9998770\pm0.0000004\;,\\\nonumber 
	R_K&=\frac{\Br(K^+\to e^+ \nu)}{\Br(K^+\to\mu^+\nu)}=
	\left(2.489\pm0.011\right)\times 10^{-5}\;,
& \Br(K^+\to\mu^+\nu)&=(63.55\pm0.11)\times 10^{-2}\;.
\end{align}
However we expect our calculation to be precise at the level of 5\% and thus our theoretical precision does not match the experimental precision. A precise treatment would require the inclusion of higher order corrections in chiral perturbation theory, which  has been done for the SM contribution in Ref.~\cite{0707.3439v1}. As there are interference terms between the SM and the new physics contribution, it is not possible to use the precise SM result directly. We do not attempt to include higher-order corrections to pion and kaon decays, but conservatively require that the predicted value taking the operator and the SM contribution into account is within 5\% of the experimental value. 
Given that the precision of these measurements is $0.3\%$ ($0.4\%$) for $R_\pi$ and $R_K$ as well as $4\times 10^{-5}\%$ ($0.17\%$) for pion (kaon) decay to a muon and a neutrino, we naively (neglecting cancellations) expect that it is possible to increase the limit on the cutoff scale from $R_\pi$, $R_K$ and Br($K^+\to \mu^+\nu$) by a factor of two. Similarly taking the experimental precision into account, it might be possible to improve the limit from Br($\pi^+\to \mu^+\nu$) by up to a factor 20.

\begin{table}[btp]
\centering
\resizebox{\linewidth}{!}{
	\begin{tabular}{llcc ccccc}
\toprule
decay & constraint &\multicolumn{2}{c}{cutoff scale $\Lambda$ [TeV]} &
\multicolumn{5}{c}{Wilson coefficients}\\
      &&$\Lambda_{\mu e, e\mu, e\tau}$ &
$\Lambda_{\tau e, \tau\mu,\mu\tau}$
		&$\Xi^{u}_{ij,uu}$ & $\Xi^{d}_{ij,dd}$  &
$\Xi^{d}_{ij,ss}$  & $\Xi^{u}_{ij,cc}$   &
$\Xi^{d}_{ij,bb}$  \\
\midrule
$R_\pi$  & $ R_\pi^{exp}\pm5\%$ &
$25-280$ &
$25-260$
    &\checkmark & \checkmark &-&-&- \\
$R_K$ &  $ R_K^{exp}\pm5\%$ &  $24-160$& $24-150$ 
    & \checkmark  &- & \checkmark &-&- \\
Br($D^+\to e^+\nu$) & $<8.8\times 10^{-6}$ & $2.8-2.9$& $2.9$ 
	&-&\checkmark &-& \checkmark & -\\
Br($D_s^+\to e^+\nu$) & $<8.3\times 10^{-5}$ & $3.2-3.3$ & $3.2-3.3$
 &-&-& \checkmark & \checkmark  &- \\
Br($B^+\to e^+\nu$) & $<9.8\times 10^{-7}$ & $2.0$ & $2.0$
 &\checkmark  &- &-&- & \checkmark \\
\midrule
Br($\pi^+\to \mu^+\nu$) & Br$^{exp}\pm5\%$ & $1.9-7.4$& $1.9-9.4$
	 & \checkmark  & \checkmark &-&-& - \\
Br($K^+\to \mu^+\nu$) & Br$^{exp}\pm5\%$ & $1.7-5.8$& $1.7-7.4$
	& \checkmark  &-& \checkmark &-& -\\
Br($D^+\to \mu^+\nu$) & $(3.82\pm 0.33)\times 10^{-4}$ & $1.1-2.7$ & $1.1-3.4$
& - & \checkmark  & - & \checkmark &- \\
Br($D_s^+\to \mu^+\nu$) & $(5.56\pm 0.25)\times 10^{-3}$ & $1.3-4.3$&$1.3-5.3$
& - & - & \checkmark  &  \checkmark & -\\
Br($B^+\to \mu^+\nu$) & $<1.0\times 10^{-6}$ & $1.9-2.7$& $1.7-3.0$
& \checkmark  & -& -& -&\checkmark \\
\midrule
Br($D^+\to \tau^+\nu$) & $<1.2\times 10^{-3}$ & $0.21-0.78$ & $0.23-0.73$
 & - &  \checkmark  & - & \checkmark &- \\
Br($D_s^+\to \tau^+\nu$) & $\left(5.54\pm0.24\right)\times 10^{-2}$ &
$0.33-1.2$ & $0.33-1.1$ 
 &  - & - & \checkmark  &  \checkmark & -\\
Br($B^+\to \tau^+\nu$) & $\left(1.14\pm0.27\right)\times 10^{-4}$ & $0.49-1.3$ &
$0.49-1.2$
  &  \checkmark  & -& -& -& \checkmark \\
\bottomrule
\end{tabular}
}
\caption{Experimental constraint on the cutoff scale $\Lambda$ [TeV] of the
	effective operators from LFV leptonic charged meson decays. The second
	column indicates the relevant experimental constraint. The third and
	fourth columns give the constraint in TeV. The index of $\Lambda$
	denotes the relevant leptons of the operator. 
	The final state charged lepton in each process is right-handed and thus corresponds to
	the second index of the Wilson coefficient. 
	Measured branching ratios are imposed at the $2\sigma$ level unless otherwise specified.
 The check marks [\checkmark] indicate the constrained operator.
}
\label{tab:CClfv}
\end{table}

The decay width for charged meson decay $M^+_{kl}=(u_k\bar d_l)$ in the limit of massless neutrinos is given by
\begin{align}
\Gamma(M^+_{kl}\rightarrow \ell_i^+ \nu)&=\frac{k_\ell}{8\pi m_M^2}\left(m_M^2- m_{\ell_i}^2\right)
\Bigg[
2G_F^2 f_M^2 m_{\ell_i}^2\left|V_{kl}\right|^2 \\\nonumber
&+\frac{m_M^2\bar
f_M^2}{4}\sum_j\left|\Xi^{Cu}_{ij,kl}-\Xi^{Cd}_{ij,kl}+\frac{y_{\ell_i}\left(y_{u_k}+y_{d_l}\right)}{m_W^2}U_{ij}^*V_{kl}^*\right|^2\\\nonumber
&-\sqrt{2}G_F m_{\ell_i}m_M \bar f_M f_M
\left(\frac{y_{\ell_i}\left(y_{u_k}+y_{d_l}\right)}{m_W^2} |V_{kl}|^2 +
\mathrm{Re}\left(\sum_j\left(\Xi^{Cu}_{ij,kl}-\Xi^{Cd}_{ij,kl}\right)U_{ij} V_{kl}\right) \right)
\Bigg]
\end{align}
with the 3-momentum $k_\ell$ defined in Eq.~\eqref{eq:kl}. The Yukawa couplings of the charged fermions are defined as $y_{u_k}\equiv m_{u_k}/v$, $y_{d_l}\equiv m_{d_l}/v$ and $y_{\ell_i}\equiv m_{\ell_i}/v$ with the vacuum expectation value $v=174$ GeV. Finally the meson decay constant $f_M$ and the scale-dependent scalar meson decay constant $\bar f_M$ are both given in App.~\ref{app:mesons}.

All results are summarised in Tab.~\ref{tab:CClfv}. The first column lists the
observable, like the ratio $R_{\pi,K}$ and the branching ratios. The second
column indicates the used experimental constraint. Note that the calculation is
limited by the theory error in case of pions and Kaons. We require the new
physics contribution to deviate from the experimental result by less than
$2\sigma$. The third and fourth column list the obtained lower limit on the cutoff scale
$\Lambda_{ij}$, where the indices indicate the two leptons of the operator. Check marks [\checkmark] in the fifth to ninth column indicate the
operators, which are constrained by the considered process. The charged lepton
in the final state of the different processes is right-handed, {\it i.e.}~the one with
the index $j$ of the Wilson coefficient. Despite our crude calculation the
strongest constraints on the cutoff scale $\Lambda$ are extracted from the
ratios $R_\pi$ and $R_K$, which could be improved with a more precise
calculation. However they are outperformed by $\mu$-$e$ conversion in nuclei.

\section{LHC Search}
\label{sec:lhc}
At colliders, the four fermion interaction $\bar \ell_{Li} \ell_{Rj} \bar q_k
q_k$ can lead to the charged lepton flavour violating processes\footnote{The
	other four fermion interaction with a neutrino will lead to the
	signature of a mono-lepton with missing energy, which has a large SM
background from $W$-boson production and it will thus not lead to competitive limits.
Hence we do not consider it for the LHC study.},
\begin{equation}
pp\rightarrow \ell_i \ell_j +  jets\; . 
\end{equation}    
\begin{figure}[bt]
\centering
\begin{subfigure}{0.25\linewidth}
\centering
\begin{tikzpicture}[node distance=1cm and 1cm]
\coordinate[vertex] (v);
\coordinate[above left = of v, label=left:$q$] (f1);
\coordinate[below left = of v, label=left:$q$] (f2);
\coordinate[above right = of v, label=right:$\ell_i$] (f3);
\coordinate[below right = of v, label=right:$\ell_j$] (f4);
\draw[fermionnoarrow] (f1) -- (v);
\draw[fermionnoarrow] (f2) -- (v);
\draw[fermionnoarrow] (f3) -- (v);
\draw[fermionnoarrow] (f4) -- (v);
\end{tikzpicture}
\caption{}\label{fig:LHCtree}
\end{subfigure}
\hfill
\begin{subfigure}{0.24\linewidth}
\centering
\begin{tikzpicture}[node distance=1cm and 1cm]
\coordinate[vertex] (v);
\coordinate[above left = of v, label=left:$q$] (f1);
\coordinate[below left = of v, label=left:$q$] (f2);
\coordinate[above right = of v, label=right:$\ell_i$] (f3);
\coordinate[below right = of v, label=right:$\ell_j$] (f4);
\coordinate[above = of f3, label=right:$g$] (g);
\draw[gluon] (g) -- ($(f1)!0.5!(v)$);
\draw[fermionnoarrow] (f1) -- (v);
\draw[fermionnoarrow] (f2) -- (v);
\draw[fermionnoarrow] (f3) -- (v);
\draw[fermionnoarrow] (f4) -- (v);
\end{tikzpicture}
\caption{}\label{fig:LHCgluon}
\end{subfigure}
\hfill
\begin{subfigure}{0.24\linewidth}
\centering
\begin{tikzpicture}[node distance=1cm and 1cm]
\coordinate[vertex] (v1);
\coordinate[vertex,below = of v1] (v2);
\coordinate[left = of v1, label=left:$g$] (g);
\coordinate[left = of v2, label=left:$q$] (f1);
\coordinate[above right = of v1, label=right:$q$] (f2);
\coordinate[above right = of v2, label=right:$\ell_i$] (f3);
\coordinate[below right = of v2, label=right:$\ell_j$] (f4);
\draw[fermionnoarrow] (f1) -- (v2);
\draw[fermionnoarrow] (f2) -- (v1);
\draw[fermionnoarrow] (f3) -- (v2);
\draw[fermionnoarrow] (f4) -- (v2);
\draw[fermionnoarrow] (v1) -- (v2) node[midway,label=left:$q$] {};
\draw[gluon] (g) -- (v1);
\end{tikzpicture}
\caption{}\label{fig:LHCujet1}
\end{subfigure}
\hfill
\begin{subfigure}{0.24\linewidth}
\centering
\begin{tikzpicture}[node distance=1cm and 1cm]
\coordinate[vertex] (v1);
\coordinate[vertex,right = of v1] (v2);
\coordinate[above left = of v1, label=left:$g$] (g);
\coordinate[below left = of v1, label=left:$q$] (f1);
\coordinate[above right = of v2, label=right:$q$] (f2);
\coordinate[right = of v2, label=right:$\ell_i$] (f3);
\coordinate[below right = of v2, label=right:$\ell_j$] (f4);
\draw[fermionnoarrow] (f1) -- (v1);
\draw[fermionnoarrow] (f2) -- (v2);
\draw[fermionnoarrow] (f3) -- (v2);
\draw[fermionnoarrow] (f4) -- (v2);
\draw[fermionnoarrow] (v1) -- (v2) node[midway,label=below:$q$] {};
\draw[gluon] (g) -- (v1);
\end{tikzpicture}
\caption{}\label{fig:LHCujet2}
\end{subfigure}
\caption{Signatures at hadron colliders.}
\label{fig:signatures}
\end{figure}
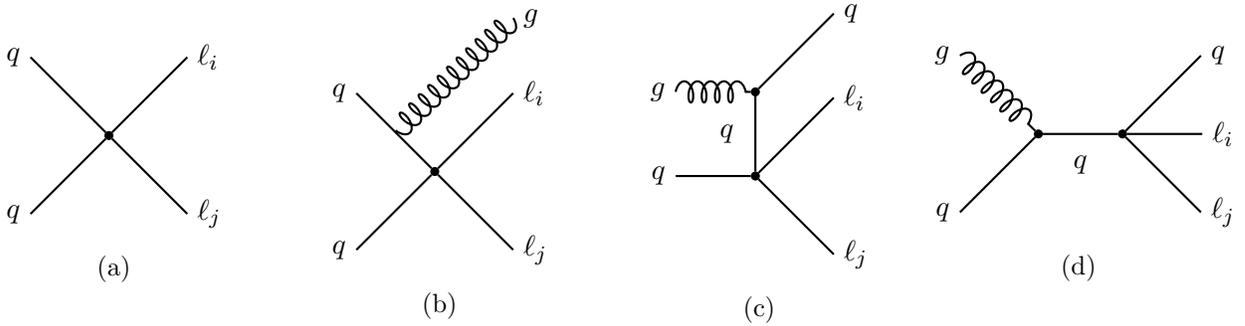 
We show the Feynman diagrams contributing to this process up to one jet in Fig.~\ref{fig:signatures},
including the leading order contribution in Fig.~\ref{fig:LHCtree} and the
next-to-leading order contributions in
Figs.~\ref{fig:LHCgluon}-\ref{fig:LHCujet2}. 

CDF and D0 reported on their search for $e\mu$ final states from $s$-channel heavy resonance decays in Refs.~\cite{Abulencia:2006xm,Abazov:2010km}.
There are also rich studies about charged lepton flavour violating processes at the LHC.
ATLAS has searched for $Z\to e\mu$ in Ref.~\cite{Aad:2014bca}.
Similarly LFV Higgs decay has also been studied in Refs.~\cite{Khachatryan:2015kon, Aad:2015gha}.   
Both ATLAS and CMS have performed search for heavy resonances decay to $e\mu$ in Refs.~\cite{Aad:2011kta, CMS-PAS-EXO-13-002}.
ATLAS has also expanded their search to include $e\mu$, $e\tau$ and $\mu\tau$ in \cite{Aad:2015pfa}.
These analyses examined the $e\mu$, $e\tau$ or $\mu\tau $ invariant mass spectrum for the presence of a heavy particle.
They found no evidence of new physics and gave model-dependent limits on the mass of the heavy resonances for given couplings.
All these searches looked for LFV processes {\it inclusively}, i.e.~including extra jets.  
In Ref.~\cite{Aad:2012yw} ATLAS searched {\it exclusively} for final states with a LFV $e\mu$ pair and zero jet 
for $t$-channel $\tilde{t}$ exchange.
Note that in most analyses well-defined and properly reconstructed jets have $p_T\gtrsim 30 \; \rm{GeV}$.

We will take the most up-to-date {\it inclusive} and {\it exclusive} analyses for a pair of oppositely charged flavour off-diagonal leptons, 
i.e.~the 8 TeV search for $e\mu$, $e\tau$ and $\mu\tau$ with $20.3$ fb$^{-1}$ integrated luminosity in Ref.~\cite{Aad:2015pfa} and the 7 TeV search for $e\mu$ with $2.08$ fb$^{-1}$ integrated luminosity in Ref.~\cite{Aad:2012yw}.
The searches have quite distinctive SM background because of the requirement on jets, which will be elaborated in Sec.~\ref{subsec:sigandbkg}. 
With Monte Carlo simulation and the aid of hepdata, we will recast both searches and extract the LHC limits for the 
effective operators chosen in this work. 

\begin{figure}[bt]
\centering
\includegraphics[width=0.7\linewidth]{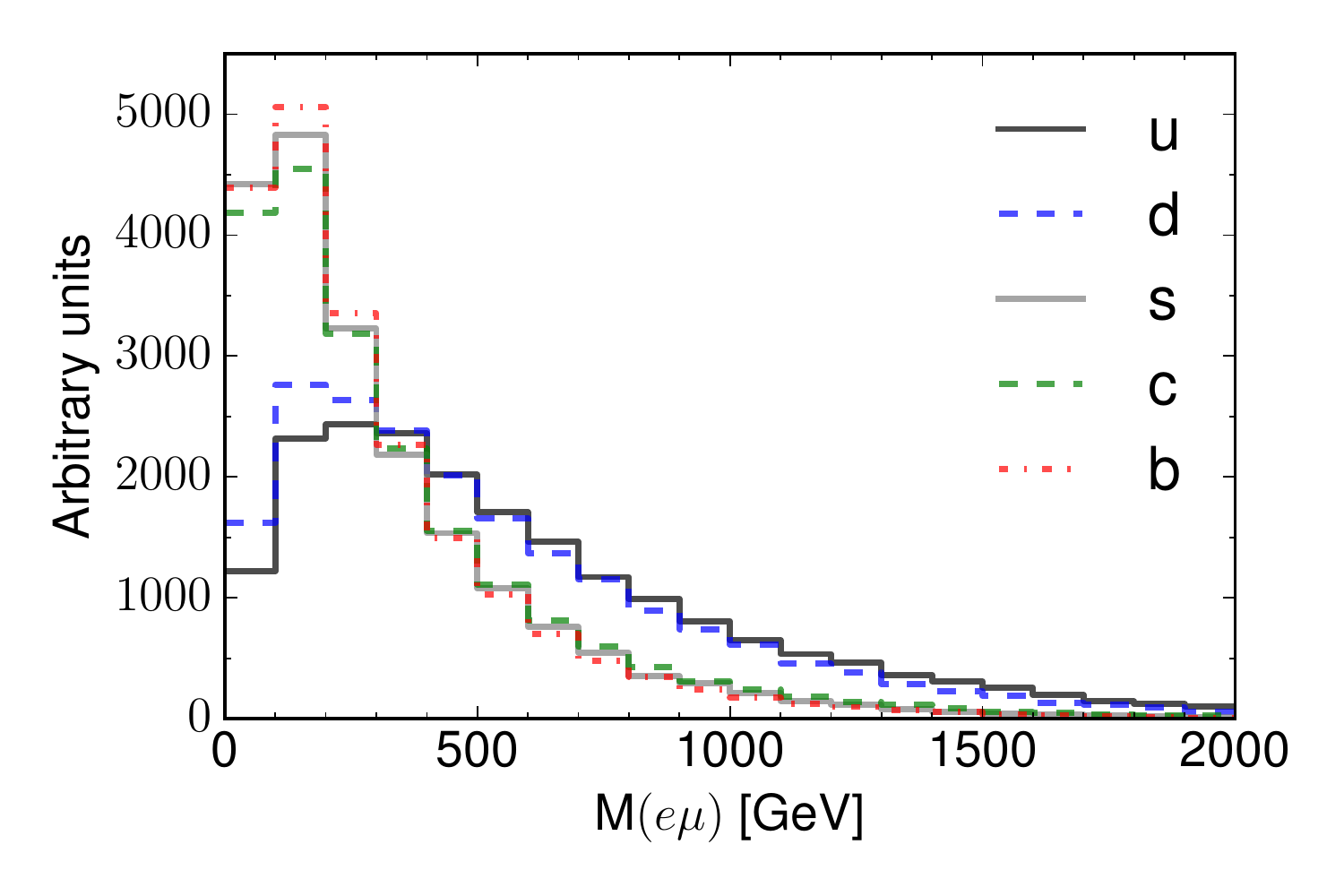}
\caption{The distribution of the $e\mu$ invariant mass at parton level at $\sqrt{s}=7\; \rm{TeV}$. The production cross section of each operator has been normalised to the same value. }
\label{fig:qqemu}
\end{figure}

Before we move on to the details of the simulation, we want to stress that the LHC limits depend on the quark flavour in a
not-so-trivial manner. Because of the parton distribution functions, the $p_T$ distribution and the invariant mass distribution 
of the lepton pairs in the final states are also different for operators with different quark flavours 
even if the total production cross sections at the LHC were the same.   
As an example, we plot invariant mass distribution of $e\mu$ final states in a $pp$ collider at $\sqrt{s}=7\; \rm{TeV}$ in Fig.~\ref{fig:qqemu}. It is easy to see that the distributions are quite similar for the two lighter flavours $u$ and $d$, plotted with a black solid and blue dashed line respectively, and for the heavier ones $s$, $c$ and $b$ shown with a gray solid, green dashed, and red dot-dashed line respectively.

\subsection{Signal and Background}
\label{subsec:sigandbkg}
The signals for the 7 TeV $e\mu$ exclusive analysis dominantly come from the tree-level process in Fig.~\ref{fig:LHCtree}.
Since the exclusive search rejects any events with a well-fined jet, we neglect all  next-to-leading order contributions for the signals.
The relevant operators are implemented in {\it FeynRules 2.0}~\cite{Alloul:2013bka} to generate output model files in UFO format. 
The signal events are generated in {\it MadGraph 5}~\cite{Alwall:2011uj} 
at leading order with parton distribution function nn23lo1.
The parton level events are subsequently piped to {\it PYTHIA 8.2}~\cite{Sjostrand:2007gs} for showering and hadronization.
The detector effects are simulated using {\it Delphes 3}~\cite{deFavereau:2013fsa}.     
For operators  with $e\tau$ and $\mu\tau$, the $\tau$-lepton could also decay leptonically and gives an $e\mu$ final state.
However, these processes are suppressed by the leptonic branching ratios of $\tau$ and lead to really poor limits. 
Thus we will only consider operators with $e\mu$ for the 7 TeV analysis.  

Similar to the 7 TeV search, the signal of the inclusive search at 8 TeV also comes mainly from the tree level diagram in Fig.~\ref{fig:LHCtree}. The next-to-leading order contribution can result in a $K$-factor.
Assuming a uniform $K$-factor, the lower limit on the UV cutoff will be scaled up by $K^{\frac{1}{4}}$, 
which only improves the limits  by a few percent.
So for the 8 TeV analysis in this work, we will take the leading order contribution from the tree level diagram and 
assume a unity $K$-factor for simplicity. 
The signal samples are  generated with the same tool chain.                

The major SM processes that can lead to $e\mu$ final states include 
$t\bar{t}$, $WW$, and $Z/\gamma^*\to \tau\tau$.
The $t\bar{t}$ pair decays to $e\mu$ via $W$ bosons and is always accompanied with two hard $b$-jets.
The other two channels, $WW$ and $\tau\tau$, give rise to a LFV lepton pair through leptonic $W$ and $\tau$ decay,
which usually have large missing transverse energy, $E_T^{miss}$, because of the neutrinos in the final states. 
So in the 7 TeV analysis, these background events are quite efficiently eliminated by the 
selection requirements for zero jet and small $E_T^{miss}$. 
Because jets can be misidentified as leptons, 
$W/Z$ plus jets and multi-jets also contribute.
This type of background is denoted as fake background and estimated from data at the ATLAS search.
Other subdominant background includes $WZ/ZZ$, single top and $W/Z+\gamma$.
For both 7 TeV and 8 TeV analysis, $WW$, $Z/\gamma^*\to\tau\tau$ and $t\bar{t}$ 
make up around $90\%$ of the background. 
Because of the strict selection rules, the background for the 7 TeV analysis is much cleaner than the 8 TeV one.
This ensures a good limit even with a much lower integrated luminosity at 7 TeV.   
For the $e\tau$ and $\mu\tau$ search at the 8 TeV, the background is dominated by the Drell-Yan process $Z/\gamma^*\to \mu\mu/\tau\tau$ and the fake background. The contribution from the fake background can be as much as $50\%$.    

We use {\it MadGraph 5} at NLO to generate the background sample for $WW$, $Z/\gamma^*\to \tau\tau$ and $t\bar{t}$, 
where showering and hadronization is handled with {\it Herwig 6}~\cite{Corcella:2000bw}.
Detector effects are simulated with {\it Delphes 3}. Our simulated background samples agree with the experimental analysis.
However, the simulation and the estimation of the fake background requires the analysis on the actual experimental data,
which is way beyond the scope of this work.
Therefore, we will use the experimental measurements to extract the limits at both 7 TeV and 8 TeV.

For the 14 TeV LHC run, we will only try to project the limit on the operators with $e\mu$ 
because of the non-negligible fake background for other final states. 
Of the two searching strategies, we will choose the one in the 7 TeV analysis, which gives a much cleaner background and 
thus a better limit for the same dataset.
So we will make use of the tool chain described here for the $WW$ and $Z/\gamma^*\to \tau\tau$ background estimate.
We assume the contribution from the fake background in the selected sample at 14 TeV
will be slightly less than that from $Z/\gamma^*\to \tau\tau$ as in the 7 TeV analysis. In our simulation we will consider an integrated luminosity of $300$ fb$^{-1}$.

\subsection{Event Selection}
For the 7 TeV and 8 TeV search, we take  the same selection requirements as in the ATLAS analysis. 
The event selection requires a pair of oppositely charged leptons.
Electrons should have $E_T>25 \; \rm{GeV}$ and satisfy a set of stringent identification requirements referred as {\it tight}.
We implement the {\it tight} identification through the electron efficiency in {\it Delphes 3} as in Ref.~\cite{Drees:2013wra}.
Electrons are rejected if they lie outside the pseudorapidity regions $\left|\eta\right|<1.37$ or $1.52<\left|\eta\right|<2.47$.
Similarly muons are required to have $p_T>25\; \rm{GeV}$ and $\left|\eta\right|<2.4$.
Tau candidates should also have $E_T > 25 \; \rm{GeV}$ and lie in the proper pseudo-rapidity range $\left|\eta\right|<2.47$
and $\left|\eta\right|>0.03$. 
In addition, we implement the lepton isolation requirements: 
the scalar sum of the track $p_T$ within a cone of $\Delta R=0.2 (0.4)$ around the lepton is less than $10\%$ ($6\%$) 
of the lepton's $p_T$ for the 7 (8) TeV search; 
similarly the sum of $E_T$ within the cone of $\Delta R= 0.2$ is less than $15\%$ ($6\%$) of the lepton's $E_T$ for the 7 (8) TeV search.
Jets are reconstructed using the anti-$k_t$ algorithm with a radius parameter of 0.4.
In the 7 TeV search any events with jets that have $p_T>30\; \rm{GeV}$ or $E_T^{miss}<25 \;\rm{GeV}$ are rejected. 
Additionally the invariant mass of the lepton pair should be bigger than 100 (200) GeV 
and the azimuthal angle difference between them should be bigger than 3 (2.7) for the 7 (8) TeV search.

For the 14 TeV projection, we will impose the following cuts on lepton transverse momentum, $p_T$, and the total missing transverse energy:
$p_T(\ell)>300\; \rm{GeV}$ and $E_T^{miss}<20\; \rm{GeV}$, in addition to the same cuts on the azimuthal angle $\Delta \phi(e, \mu)>3.0$ and the lepton identification and isolation requirements. 
After the selection, the only SM background is from $WW$, while $\tau\tau$ contribution drops much faster with increasing dilepton invariant mass.
With the assumption that the contribution from fake background is less than that from $\tau\tau$, we can also neglect the fake background in the 
14 TeV projection.

\subsection{Limit Setting and Results}
We use maximum likelihood estimator for limit setting at the 7 and 8 TeV searches. 
The observed invariant mass distributions of the $e\mu$ pair as well as the SM background for the 7 and 8 TeV analyses 
are taken from hepdata.
The likelihood function for each bin is defined as 
\begin{equation}
\mathcal{L}_i(\mu,\tilde{\theta_i}| n_i) = \mathcal{P}(n_i|\mu \; s_i + b_i) \mathcal{G}(\tilde{\theta_i}, 0, 1),
\end{equation} 
where $\mathcal{P}$ and $\mathcal{G}$ are Poisson and Gaussian functions.
$s_i$, $b_i$ and $n_i$ are the predicted signal, SM background and the observed events in the $i$-th bin.
The parameter $\mu$ is the signal strength and $\tilde{\theta}_i$ is the nuisance parameter. 
The total likelihood function is the product of $\mathcal{L}_i$ in each bin.
This limit setting method is tested with the hypothesis as described in the 7 and 8 TeV ATLAS analyses 
and the results agree within percent level.
\begin{table}[bt]
\centering
\begin{tabular}{c | c c c|  c c }
\toprule
\backslashbox{$\bar q q$}{$\bar \ell_i \ell_j$}      & \multicolumn{3}{c}{$\bar e\mu$} &      $\bar e\tau$ &  $\bar \mu\tau$ \\
&   7 TeV & 8 TeV & 14 TeV &   8 TeV &   8 TeV\\
\midrule
$\bar uu$  & 2.6  & 2.9 & 8.9 &  2.4     & 2.2 \\
$\bar dd$  & 2.3  & 2.3 & 8.0 &  2.1     & 1.9\\
$\bar ss$  & 1.1  & 1.4 & 4.0 &  0.95    & 0.88 \\
$\bar cc$  & 0.97 & 1.3 & 3.6 &  0.82    & 0.78 \\
$\bar bb$  & 0.74 & 1.0 & 2.7 &  0.63    & 0.61 \\
\bottomrule
\end{tabular}
\caption{ Constraints from the LHC searches on the cutoff scale $\Lambda$ [TeV] at the 7 and 8 TeV search. The 14 TeV projection is also listed for the $e\mu$ final state. }\label{tab:Constraints}
\end{table}

For the 14 TeV projection, we will perform the same limit setting procedure with the binned invariant $e\mu$ mass 
spectrum from 600 GeV to 1 TeV with bin width of 50 GeV as well as an over-flow bin.  
For the 14 TeV projection, we will estimate the experimental reach simply with 
\begin{equation}
\rm{Significance} = \frac{S}{\sqrt{S+\left(\Delta S\right)^2 + \left(\Delta B\right)^2 }},
\end{equation}    
where $S$ and $B$ denote the number of signal and background events. $\Delta S$ and $\Delta B$ parameterize the systematic 
uncertainties, $\Delta S = 10\% S$ and $\Delta B= 10\% B$.     

We list the current limits and future projection at the 14 TeV run in Tab.~\ref{tab:Constraints}.
Note that the 8 TeV search does not result in much better limits even with a higher beam energy and 10 times more data than the 7 TeV one,
simply due to the large background. The limits for $e\tau$ and $\mu\tau$ are both weaker than the $e\mu$ channel at 8 TeV 
as a result of the low $\tau$-tagging rate and higher fake background.

\section{Discussion}
\label{sec:discussion}
\begin{figure}[bt]\centering
\includegraphics[width=\linewidth]{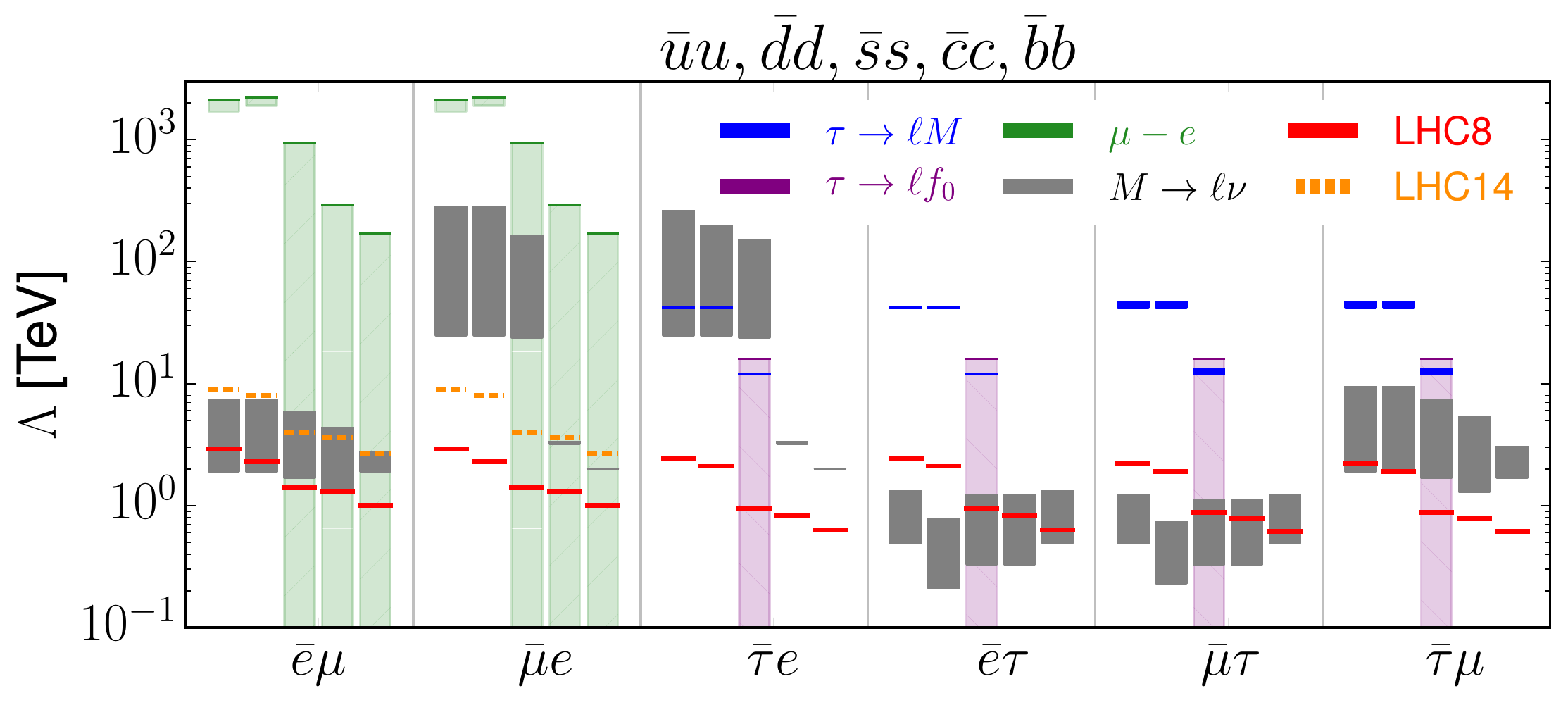}
	\caption{Summary plots of most stringent limits from precision experiments and the LHC. See the text for a detailed explanation. }
\label{fig:summary}
\end{figure}

In Fig.~\ref{fig:summary} we compare the most stringent constraints from
precision experiments and the LHC.
If the constraint depends on a free parameter like the phase of the Wilson
coefficient or a mixing angle, we show the possible constraints in a band and include
the second-most stringent constraint as well. For completeness, we kept the
constraints from $R_\pi$ and $R_K$, since they suffer from their theoretical
uncertainty and can be further improved with a more detailed calculation by a
factor of a few. Operators with different lepton
combinations are separated by a gray vertical line. The
figure shows the limits on operators with the quarks $u, d, s, c, b$ ordered
from left to right in each of the blocks. The current 8 TeV LHC constraints are
denoted by a solid red line and the
future (14 TeV) sensitivity by a dashed orange line. The constraints from $\mu$-$e$ conversion
are shown in green indicating the range between direct nuclear mediation and
meson exchange mediation. Blue lines indicate limits from $\tau$-decays to
a charged lepton and a neutral pseudoscalar mesons besides $f_0$, which is shown
in purple. It depends on the undetermined mixing
angle $\varphi_m$ between the different quark compositions. Finally, constraints from
leptonic charged meson decays are shown in gray. The limits from leptonic
neutral meson decays were generally weaker than the presented limits and are
thus not included in the figure. 

For operators with $e\mu$ across all quark flavours, the limit from $\mu$-$e$ conversion in nuclei clearly outperforms any other limit.
Even its current limit on $\Lambda$ assuming direct nuclear mediation 
is two orders of magnitude higher than that from the 14 TeV projection of the LHC. The limits certainly will be further improved by the two proposed experiments, Mu2E~\cite{Carey:2008zz,Kutschke:2011ux} at FNAL and COMET~\cite{Hungerford:2009zz,Cui:2009zz} at J-PARC, which aim to improve the sensitivity of $\mu$-$e$ conversion in $^{48}$Ti down to $10^{-16}$, and possibly even to $10^{-18}$ in a future proposed experiment PRISM/PRIME~\cite{Hungerford:2009zz,Cui:2009zz}.
For $e\tau$ and $\mu\tau$, pion and Kaon decays and semi-leptonic  $\tau$ decays places the strongest
constraints for the light quark flavours. The limits from $\tau$ decays will be further improved at the Belle-II experiment~\cite{Aushev:2010bq}: Belle-II aims to increase the sensitivity on the branching fraction by two orders of magnitude. Note that constraints from precision measurements for $\bar{u}u$ and $\bar{d}d$ are quite similar, which can be easily explained with the isospin symmetry. 

However, the constraints on the operators with heavy quark flavours are
generally weaker. That is exactly where the LHC comes into play.
With the 8 TeV LHC search, the collider limit is competitive with constraints
from $\tau$-decays and charged meson decays for operators with
$\bar{c}c$ and $\bar b b$ and a right-handed $\tau$-lepton. In particular it
does not depend on the phase of the Wilson coefficient. The limits from charged
meson decays feature an interference with the SM contribution and thus depend on
the phase of the Wilson coefficient. Using the 8 TeV LHC data, we set limits of
600-800 GeV on the cutoff scale $\Lambda$ for LFV operators with right-handed $\tau$ leptons and we expect that those limits can be further improved with more integrated luminosity, similar to projected sensitivity for the $e\mu$ channel at $14$ TeV with 300 fb$^{-1}$. In case of the $e\mu$ channel we find that the sensitivity of the LHC can be improved by a factor of $2.7-3.5$.

Finally we want to address the validity of the effective operator descriptions at the LHC. When the momentum transfer $Q_{tr}$ 
in the interaction is comparable to the heavy mediator of mass $M$ in the UV completion of the effective theory, 
the effective operator description is no longer a good approximation. The UV cutoff scale $\Lambda$ is related to the mass of 
the heavy mediator with 
\begin{equation}
M = \Lambda \sqrt{g_q g_l},
\end{equation}   
where $g_{q}$ ($g_l$) denote the couplings between the heavy mediator and the quarks (leptons).
To properly preserve the validity of the effective operators, 
a procedure referred as truncation can be conducted when $Q_{tr}>M$, i.e. the event is discarded~\cite{Busoni:2013lha}.    
So if we take the optimistic limit $g_{q}=g_{l}=4\pi$, the heavy mediator mass should be at least $\gtrsim 7.7$ TeV for the 8 TeV search
and the results are surely valid.
Even with relatively conservative option $g_{q}=g_{l}=1$, the LHC limit is still quite sound,
because the LHC  analyses we use in this work rely mostly on events with smaller momentum transfer. 
Moreover, if very small values of $g_q$ and $g_l$ are chosen, we are bound to return to the UV completions,
which is a completely different type of study not meant to be contained in this work. 
\section{Conclusions}
\label{sec:con}

From the comprehensive case study in this work, we see that precision
measurements and the LHC study are indeed complementary. Which experiment gives
the best reach depends on both the quark flavour and the lepton pair in the
operator. For light quarks $u$, $d$ and $s$, precision measurements clearly
outperform the LHC irrespective of the charged lepton flavour. However, the LHC
becomes competitive for heavier quarks, $c$ and $b$, and there is an interesting
interplay between the two approaches to obtain limits on LFV operators with two
quarks and two leptons. Operators with $e\mu$ are still highly constrained by
precision measurements, particularly $\mu$-$e$ conversion in nuclei, but the LHC
competes for LFV operators with right-handed $\tau$ leptons and can set limits
independent of the phase of the Wilson coefficient. We set a lower limit of 600-800 GeV on the cutoff scale of all these operators.

In this study we restricted ourselves to scalar operators and did not consider operators with top quarks. For other Lorentz structures we expect similar limits from the LHC, but the limits from the precision experiments have to be reevaluated. In case of top quarks, there are no direct limits from precision experiments, although operator mixing will lead to some constraint. We expect that a similar analysis of the collider phenomenology can set new interesting limits in addition to constraints from flavour violating top decays. Finally we only considered non-resonant searches and did not consider possible underlying UV completions. A complementary study of simplified models, where the operators are opened up, might lead to more stringent, although model-dependent, limits.

\section*{Acknowledgements}
We thank Martin Holthausen for collaboration during the initial stages of this project and Tong Li, Fei Gao, Lei Wu, Aldo Saavedra and Bruce Yabsley for useful discussions. This work was supported in part by the Australian Research Council. We acknowledge the use of \texttt{matplotlib}~\cite{Hunter:2007} and \texttt{ipython}~\cite{PER-GRA:2007}.

\appendix

\section{Mesons}\label{app:mesons}
The quark bilinear in the operator we choose to study determines that the only mesons involved in the LFV processes are either neutral scalars or pseudoscalars.  
The decay constants $f_M$ for scalar (S) and pseudo-scalar (P) mesons are defined as
\begin{table}[tb]
\centering
\begin{tabular}{lccccc}
\toprule
Meson & $m_M$/MeV & $\tau_M$/s & $\Gamma_M$/MeV & $f_M$/MeV & $\bar f_M$/MeV \\
\midrule
$\pi^0$ &  134.9766  & $8.52\times 10^{-17}$ & $7.725\times 10^{-6}$ & $130.41$ & $2500$ \\
$\eta$  & 547.862   & &1.31   &  $\star$ & $\star$  \\
$\eta^\prime$ & 957.78 && 0.198 & $\star$& $\star$ \\
\midrule
$K^0_L$  & 497.614   &$5.116\times 10^{-8}$ & $1.287\times 10^{-14} $  & $156.2$& $790$ \\
$K^0_S$  & 497.614   &$8.954\times 10^{-11}$ & $7.351\times 10^{-12} $  & $156.2$& $790$ \\
$D^0$ & 1864.84   & $4.101\times 10^{-13}$ &$1.605\times 10^{-9}$ & 204.6& 300\\
$B^0$  & 5279.58   & $1519\times 10^{-15}$ & $4.333\times 10^{-10}$ &190.6 & 240\\
\midrule
$\pi^+$ &  139.57018  &$2.6033\times 10^{-8}$ &  $2.5284\times 10^{-14}$ & 130.41& 2600\\
$K^+$  &  493.677  &$1.2380\times 10^{-8}$  &$5.3167\times 10^{-14}$ &156.2& 780\\
$D^+$  & 1869.61   &$1.040\times 10^{-12}$ &  $6.329\times 10^{-10}$& 204.6& 300 \\
$D^+_s$ & 1968.30  &$5.00\times 10^{-13} $ & $1.32\times 10^{-9}$ & 257.5& 370\\
$B^+$ &  5279.26    & $1638\times 10^{-15}$ &  $4.018\times 10^{-10}$ &190.6& 240\\
\midrule
$f_0(980)$ & 990  & & 40-100 & $\star$ & $\star$ \\
\bottomrule
\end{tabular}

\caption{Relevant data for the scalar and pseudoscalar mesons studied in this work. Besides the last meson $f_0(980)$ which has $J^{PC}=0^{++}$, all mesons are pseudoscalar mesons with $J^{PC}=0^{-+}$ according to the quark model assignment. We list the decay constants for most mesons assuming isospin symmetry to relate the decay constants of charged mesons with the corresponding neutral meson.
$\star$ Please refer to the text for the decay constants of $\eta^{(\prime)}$ and $f_0(980)$.
}
\label{tab:mesons}
\end{table} 
\begin{subequations}
\label{eq:decayconstants}
\begin{align}
 \left\langle 0|\bar q^i\gamma^\mu q^j|S(p)\right\rangle &=   f_S p^\mu\;, &
 \left\langle 0|\bar q^i q^j|S(p)\right\rangle & = 
m_S \bar{f}_S\;, \label{eqn:meson1}\\
 \left\langle 0|\bar q^i\gamma^\mu\gamma_5q^j|P(p)\right\rangle &=f_P p^\mu\;,  &
 \left\langle 0|\bar q^i\gamma_5q^j|P(p)\right\rangle &=  m_P \bar{f}_P  =m_P  f_P\frac{m_P}{m_q^i+m_q^j}  \label{eqn:meson2}\;.
\end{align}
\end{subequations}
The scale-dependent scalar decay constants $\bar f_M$ are related to the decay constants $f_M$ via the equations of motion. 

We use the experimental values (where available) for the pseudoscalar decay constants and the quark masses in Ref.~\cite{Agashe:2014kda}
\begin{align}
\bar m&=\frac{m_u+m_d}{2} = 3.5^{+0.7}_{-0.2}\;\mathrm{MeV}\;,&
m_c&= (1.275\pm0.025)\;\mathrm{GeV}\;, \\\nonumber
m_s&= (95\pm5)\;\mathrm{MeV}\;, &
m_b&= (4.18\pm0.03)\;\mathrm{GeV}
\end{align}
to obtain the scale-dependent scalar decay constants. All decay constants are listen in Tab.~\ref{tab:mesons} except for the states $\eta^{(\prime)}$ and $f_0(980)$, where the decay constants depend on a mixing angle.

The pseudo-scalars $\eta$ and $\eta^\prime$ mix with each other and are a mixture of $\ket{s \bar s}$ and the isospin singlet $\ket{q\bar q}\equiv\left(\ket{u\bar u}+\ket{d\bar d}\right)/\sqrt{2}$ and their decay constants can be parameterised in terms of two decay constants $f_{q,s}$ and two mixing angles $\phi_{q,s}$
\begin{equation}
\begin{pmatrix}
f_\eta^q & f_\eta^s\\
f_{\eta^\prime}^q & f_{\eta^\prime}^s\\
\end{pmatrix}
\equiv 
\begin{pmatrix}
f_q\cos\phi_q & -f_s \sin\phi_s\\
f_q \sin\phi_q & f_s \cos\phi_s\\
\end{pmatrix}\;.
\end{equation}
In the FKS formalism~\cite{Feldmann:1998vh,Feldmann:1998sh,Feldmann:1999uf}, the mixing angles coincide $\phi_s=\phi_q\equiv\phi$ and glueball admixtures are neglected. The vector decay constants $f_{q,s}$ and the mixing angle $\phi$ are given by~\cite{Feldmann:1998vh,Feldmann:1999uf}
\begin{align}\label{eq:fqfs}
f_q&=(1.07\pm0.02)f_\pi\;, &
f_s&=(1.34\pm0.06)f_\pi\;, &
\phi&=(39.3\pm1.0)^\circ\;.
\end{align}
The corresponding vector decay constant for the $\eta$ and $\eta^\prime$ meson are
\begin{align}
\bar f^q_\eta&=f_q \cos\phi \simeq 110\,\mathrm{MeV}\;, &
\bar f^q_{\eta^\prime}&=f_q \sin\phi \simeq 88\,\mathrm{MeV}\;, \\\nonumber
\bar f^s_\eta&=-f_s \sin\phi \simeq -110 \, \mathrm{MeV}\;,&
\bar f^s_{\eta^\prime}&=f_s \cos\phi \simeq 130\, \mathrm{MeV}\;.
\end{align}
The meson masses given in Tab.~\ref{tab:mesons} the scalar decay constants are thus
\begin{align}
\bar f^q_\eta&=f_q \cos\phi \frac{m_\eta}{2\bar m}\simeq 8400\,\mathrm{MeV} \;,&
\bar f^q_{\eta^\prime}&=f_q \sin\phi \frac{m_{\eta^\prime}}{2\bar m}\simeq12000\,\mathrm{MeV}\;, \\
\bar f^s_\eta&=-f_s \sin\phi \frac{m_\eta}{2m_s}\simeq - 320 \, \mathrm{MeV}\;,&
\bar f^s_{\eta^\prime}&=f_s \cos\phi \frac{m_{\eta^\prime}}{2 m_s}\simeq 680\, \mathrm{MeV}\;.
\end{align}
Finally, in the case of $f_0(980)$ with mass $m_{f_0(980)}=990\pm 20 \; \rm{MeV}$~\cite{Agashe:2014kda} the scale-dependent decay constant $\bar f_M$ is given by~\cite{Cheng:2013fba}
\begin{equation}
\bar{f}_{f_0(980)}=370\pm 20 \; \rm{MeV}\;.
\end{equation}
In the simple quark picture $f_0(980)$ together with $f_0(500)$ are a mixture of $\ket{s\bar s}$ and the isospin singlet $\ket{q\bar q}$. The exact mixing angle $\varphi_m$ between the $f_0(500)$ and the $f_0(980)$ meson is not known yet. See Ref.~\cite{Fleischer:2011au} for a list of experimental results. Note however that it is unclear whether the description in the simple quark picture is actually correct or whether the $f_0(980)$ is a multi-quark state~\cite{Agashe:2014kda}. We will assume the simple quark model and parameterize our result in terms of the mixing angle $\varphi_m$ between $\ket{s\bar s}$ and the isospin singlet $\ket{q\bar q}$
\begin{equation}\label{eq:f0Mixing}
\begin{pmatrix}
\ket{f_0(980)}\\
\ket{f_0(500)}\\
\end{pmatrix}
=
\begin{pmatrix}
\cos\varphi_m & \sin\varphi_m\\
-\sin\varphi_m & \cos\varphi_m\\
\end{pmatrix}
\begin{pmatrix}
\ket{s\bar s}\\
\ket{q\bar q}\\
\end{pmatrix}\;.
\end{equation}

\bibliography{qqll} 
\end{document}